\documentclass[12pt]{article}
\usepackage{amsmath,amssymb,amsthm,amsxtra,bbm,overpic,bm,epsfig,subfigure}
\usepackage{hyperref}
\usepackage{mathrsfs}
\usepackage{enumerate}
\usepackage{graphicx}
\usepackage{color}
\usepackage{comment}
\usepackage{epstopdf}
\usepackage{float}
\usepackage{cite}
\usepackage{upgreek,overpic,setspace,multirow,textcomp}

\usepackage{slashed,stmaryrd}
\allowdisplaybreaks[1]

\def\thefootnote{\fnsymbol{footnote}}

\usepackage{slashed,stmaryrd}

\usepackage{lscape}%
\usepackage{array}
\usepackage{booktabs}%

\begin{document}
	
	\vspace{0.2cm}
	
	\begin{center}
		{\Large\bf A Nearest-neighbor Expansion of Lepton Flavor Mixing in Powers of the $\mu$-$\tau$ Permutation Symmetry Breaking Effect}
	\end{center}
	
	\vspace{0.2cm}
	
	\begin{center}
		{\bf Jihong Huang~$^{1,2}$}\footnote{E-mail: huangjh@ihep.ac.cn}
		\\
		\vspace{0.2cm}
		{\small $^{1}$Institute of High Energy Physics, Chinese Academy of Sciences, Beijing 100049, China \\
		$^{2}$School of Physical Sciences, University of Chinese Academy of Sciences, Beijing 100049, China}
	\end{center}

	\vspace{0.5cm}
	
	\begin{abstract}
		We point out that the observed pattern of lepton flavor mixing can be well described by a proper nearest-neighbor expansion of a constant $3\times 3$ unitary matrix in powers of a small parameter characterizing the fine effect of $\mu$-$\tau$ permutation symmetry breaking. We take an example of this kind for illustration, and provide complete discussions on the usefulness in the study of leptonic CP violation and unitarity triangles in matter.
	\end{abstract}
	
	
	
	\def\thefootnote{\arabic{footnote}}
	\setcounter{footnote}{0}
	
	\newpage
	
	\section{Motivation}
	
	\label{sec:motivation}
	
\begin{figure}[h!]
		\centering
		\includegraphics[scale=0.37]{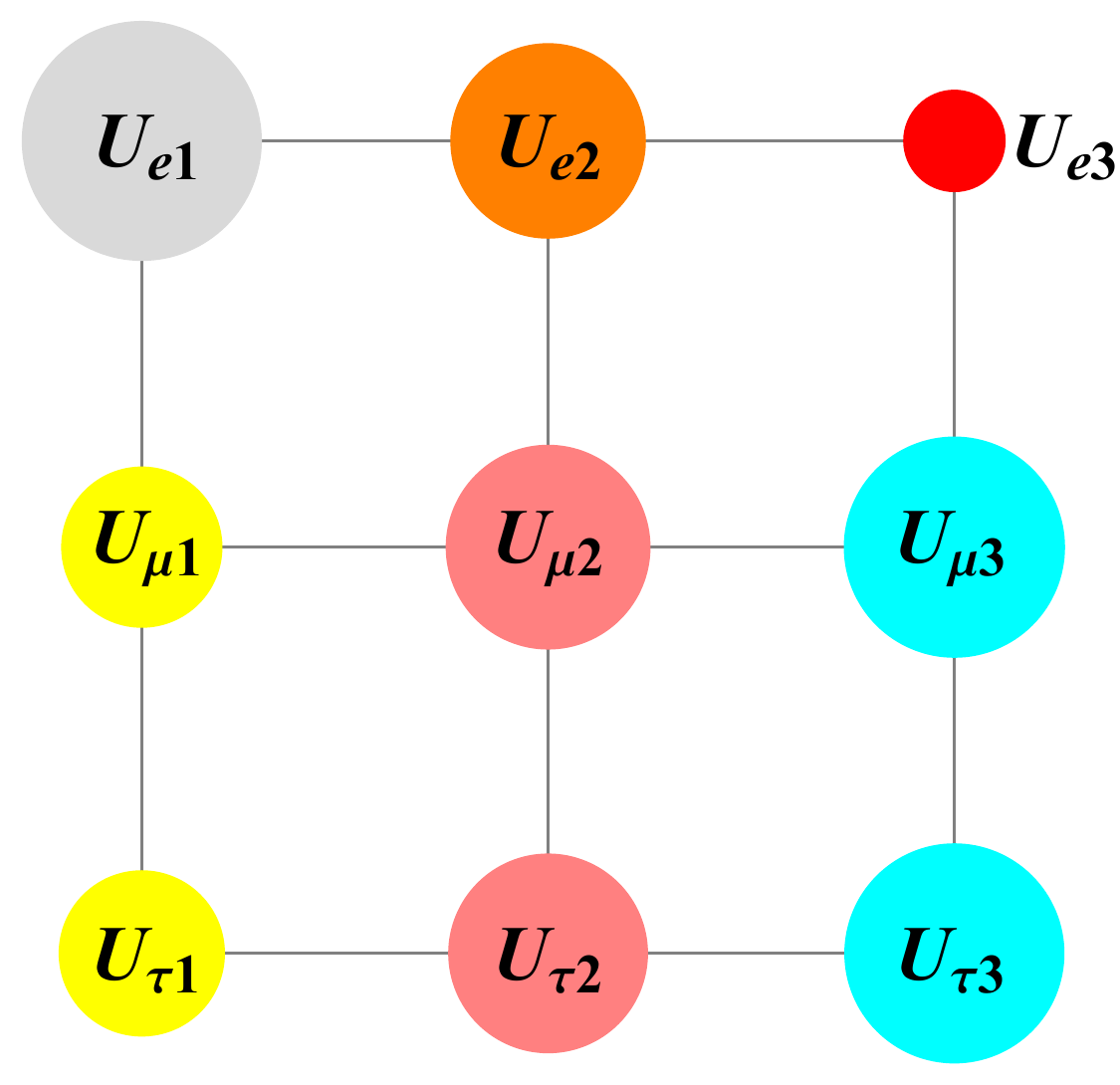}
		\caption{An illustration of the relative sizes of the PMNS matrix elements~\cite{Xing:2022uax}, where the colored area of each circle is proportional to $|U^{}_{\alpha i}|$ (for $\alpha = e, \mu, \tau$ and $i = 1, 2, 3$) on the same scaling.}
		\label{Figure1}
	\end{figure}

	The experimental discoveries of atmospheric, solar, reactor and accelerator neutrino oscillations constitute a great breakthrough in particle physics, as they clearly indicate that the neutrinos must have finite but tiny masses and the lepton flavors must be significantly mixed~\cite{ParticleDataGroup:2022pth}. Different from the Cabibbo-Kobayashi-Maskawa (CKM) quark flavor mixing matrix $V$~\cite{Cabibbo:1963yz,Kobayashi:1973fv}, which can be described as a kind of small perturbation to the identity matrix~\cite{Wolfenstein:1983yz}, the Pontecorvo-Maki-Nakagawa-Sakata (PMNS) lepton flavor mixing matrix $U$~\cite{Pontecorvo:1957cp,Maki:1962mu,Pontecorvo:1967fh} exhibits an approximate $\mu$-$\tau$ permutation symmetry in magnitude (i.e., $|U^{}_{\mu i}| \approx |U^{}_{\tau i}|$ for $i = 1, 2, 3$) as shown in Fig.~\ref{Figure1}. This observation points to $\theta^{}_{23} \approx 45^\circ$ in the standard Euler-like parametrization of $U$~\cite{ParticleDataGroup:2022pth}, the most salient feature of lepton flavor mixing which may easily arise from some simple discrete flavor symmetries~\cite{Xing:2015fdg,Xing:2022uax}. Moreover, the other two lepton flavor mixing angles are found to be $\theta^{}_{12} \approx 33.4^\circ$ and $\theta^{}_{13} \approx 8.6^\circ$ in the same parametrization of $U$~\cite{Gonzalez-Garcia:2021dve,Capozzi:2021fjo}. So the very early conjecture that the pattern of $U$ should consist of a constant symmetry-driven large-angle flavor mixing matrix $U^{}_0$ plus some small corrections characterized by the smallest flavor mixing angle and CP-violating effects~\cite{Fritzsch:1995dj,Fritzsch:1998xs,Xing:2002sw} turns out to be a widely accepted way to understand the flavor structure of lepton flavor mixing and CP violation. A newly proposed viable example of this kind is to take the tri-bimaximal flavor mixing pattern~\cite{Xing:2002sw,Harrison:2002er,He:2003rm} for $U^{}_0$ and adopt $\zeta \equiv |U^{}_{e 3}| \approx 0.149$ as the expansion parameter to parametrize the PMNS matrix $U$~\cite{Xing:2024pal}.  
	
	Note, however, that the unsuppressed size of $\theta^{}_{13}$~\cite{Fogli:2008jx,T2K:2011ypd,DayaBay:2012fng,RENO:2012mkc} {\it did} have motivated some authors to speculate that the dominant constant flavor mixing matrix $U^{}_0$ may contain all the three flavor mixing angles and even the CP-violating phases of special values (see, e.g., Refs.~\cite{Xing:2008ie,Xing:2010pn,Rodejohann:2011uz,Minkowski:2012cf} for illustration), although it is in general more difficult to derive such a pattern of $U^{}_0$ from a specific flavor symmetry model. If a symmetry-driven constant pattern of $U^{}_0$ is as close as possible to the observed pattern of $U$, then the latter can be easily obtained from a proper {\it nearest-neighbor expansion} (NNE) of the former. In this case, the perturbation to $U^{}_0$ can be sufficiently small, and hence it will be more easily attributed to a kind of flavor symmetry breaking or quantum correction effect. On the other hand, the NNE form of $U$ may help a lot to express the leading-order results of the observable quantities of lepton flavor mixing and CP violation in terms of a few simple numbers (e.g., $1/\sqrt{2}$ and $1/\sqrt{3}$) to a good degree of accuracy, and thus it is also useful from a purely phenomenological point of view. 
	
	In this work, we are going to explore the NNE approach towards a phenomenologically useful parametrization of the PMNS matrix $U$. Our main strategies are outlined below.
	\begin{itemize}
		\item We choose a typical pattern of $U^{}_0$, for the sake of illustration,
		as follows:
		\begin{eqnarray}
			\theta^{(0)}_{23} = 45^\circ \; , \quad 
			\theta^{(0)}_{12} = \arctan(1/\sqrt{2}) \approx 35.3^\circ \; , \quad 
			\theta^{(0)}_{13} = \theta^{(0)}_{23} - \theta^{(0)}_{12} \approx 9.7^\circ \;,
			\hspace{1cm}
			\nonumber 
		\end{eqnarray}
		where the values of $\theta^{(0)}_{12}$ and $\theta^{(0)}_{23}$ are fully consistent with those of the tri-bimaximal flavor mixing pattern~\cite{Xing:2002sw,Harrison:2002er,He:2003rm}, and the values of $\theta^{(0)}_{13}$ are borrowed from the constant flavor mixing ans$\rm\ddot{a}$tze proposed in Ref.~\cite{Xing:2010pn}.
		
		\item We choose the expansion parameter as $\zeta \equiv \theta^{}_{23} - \theta^{(0)}_{23} = \theta^{}_{23} - 45^\circ$, which is a meaningful measure of the small $\mu$-$\tau$ permutation symmetry breaking effect. The best-fit value of $\theta_{23}^{}$ indicates $\zeta$ is of ${\cal O}\left(10^{-2} \right)$. As its sign is sensitive to the octant of $\theta^{}_{23}$, it will be determined in the upcoming neutrino oscillation experiments. 
		
		\item We define the deviations of $\theta^{}_{12}$ and $\theta^{}_{13}$ from their corresponding values of $U^{}_0$ as $\theta^{}_{12} = \theta^{(0)}_{12} + A \zeta$ and $\theta^{}_{13} = \theta^{(0)}_{13} + B \zeta$, respectively, where $A$ and $B$ are the real parameters of ${\cal O}(1)$ or smaller. With the best-fit values of lepton mixing angles in Ref.~\cite{Gonzalez-Garcia:2021dve}, we evaluate the corresponding values of $\zeta$, $A$ and $B$ in both normal mass ordering (NO) and inverted mass ordering (IO), listed in Table.~\ref{table:zeta_A_B_9.7}
	\end{itemize}
	
	\begin{table}[h!]
		\centering
		\begin{spacing}{1.25}
			\begin{tabular}{cccc}
				\hline\hline
				& $\zeta$ & $A$ & $B$ \\ \hline
				NO ($\theta_{23}^{(0)} \approx 42.1^\circ$)	& $-0.05$ & $+0.63$ &  $+0.38$\\
				IO ($\theta_{23}^{(0)}\approx 49.0^\circ$)	& $+0.07$ & $-0.45$ & $-0.28$ \\ \hline\hline
			\end{tabular}
		\end{spacing}
		\caption{The corresponding values of $\zeta$, $A$ and $B$ in the case of normal and inverted mass ordering with $\theta_{12}^{(0)} \approx 33.45^\circ$, $\theta_{13}^{(0)} \approx 9.7^\circ$ and $\theta_{23}^{(0)} = 45^\circ$.}
		\label{table:zeta_A_B_9.7}
	\end{table}
	
	As the CP-violating phase $\delta$ in the standard parametrization of $U$ is poorly known, we keep it as a free parameter in our NNE expansion of $U^{}_0$. We leave aside the two Majorana CP phases, since they are completely unknown. When applying our parametrization to the lepton-number-violating processes such as the neutrinoless double-beta decays, one may simply put back these two phase parameters in the expression of $U$.
	
	The remaining parts of this paper are organized as follows. We expand three mixing angles and illustrate the NNE form of $U$ in Sec.~\ref{sec:expansion}, with discussions on $\mu$-$\tau$ permutation symmetry and leptonic CP violation. Further discussions on the ordering of matrix elements are also provided. In Sec.~\ref{sec:unitarity} we plot the unitary triangles and show the modification by terrestrial matter effects. The impact of non-unitarity on the mixing matrix will also be discussed. We summarize our main results in Sec.~\ref{sec:summary}. For completeness, the exact formulas of effective neutrino masses and mixing angles are listed in Appendix~\ref{app:A}.
	
	\section{The Expansion of $U$}
	\label{sec:expansion}
	
	The so-called standard parametrization of the unitary $3\times 3$ PMNS matrix $U$ advocated by the Particle Data Group (PDG) reads~\cite{ParticleDataGroup:2022pth}
	\begin{eqnarray}\label{eq:PMNS_standard}
		U = \begin{pmatrix}
			c_{12}^{} c_{13}^{} & s_{12}^{} c_{13}^{} & s_{13}^{} {\rm e}^{-{\rm i} \delta}  \\
			-s_{12}^{} c_{23}^{} - c_{12}^{} s_{13}^{} s_{23}^{} {\rm e}^{{\rm i}\delta} & c_{12}^{} c_{23}^{} - s_{12}^{} s_{13}^{} s_{23}^{} {\rm e}^{{\rm i} \delta} & c_{13}^{} s_{23}^{} \\
			s_{12}^{} s_{23}^{} - c_{12}^{} s_{13}^{} c_{23}^{} {\rm e}^{{\rm i}\delta} & -c_{12}^{} s_{23}^{} - s_{12}^{} s_{13}^{} c_{23}^{} {\rm e}^{{\rm i} \delta} & c_{13}^{} c_{23}^{} 
		\end{pmatrix} \;,
	\end{eqnarray}
	where $s_{ij}^{} \equiv \sin\theta_{ij}^{}$ and $c_{ij}^{} \equiv \cos\theta_{ij}^{}$ for $ij=12,13,23$. Two Majorana phases can always be left out and we do not involve them in this work as mentioned before. Following the strategies, the mixing angles by expanding with the parameter $\zeta$ up to ${\cal O}\left(\zeta^3\right)$ are:
	\begin{eqnarray}
		\label{eq:sin_cos_theta}
		s_{12}^{} &\approx& \frac{1}{\sqrt{3}} \left(1 + \sqrt{2} A \zeta - \frac{1}{2} A^2 \zeta^2\right)  \;, \nonumber \\
		c_{12}^{} &\approx& \frac{\sqrt{2}}{\sqrt{3}} \left(1- \frac{1}{\sqrt{2}} A\zeta - \frac{1}{2} A^2 \zeta^2\right)  \;, \nonumber \\
		s_{13}^{} &\approx& \frac{\sqrt{2}-1}{\sqrt{6}}\left[1 + \left(3+2\sqrt{2}\right) B \zeta - \frac{1}{2} B^2 \zeta^2 \right]  \;, \nonumber \\
		c_{13}^{} &\approx& \frac{\sqrt{2}+1}{\sqrt{6}}\left[1 - \left(3-2\sqrt{2}\right) B \zeta - \frac{1}{2} B^2 \zeta^2\right]  \;, \nonumber \\
		s_{23}^{} &\approx& \frac{1}{\sqrt{2}}\left(1 +  \zeta - \frac{1}{2} \zeta^2\right)  \;, \nonumber \\
		c_{23}^{} &\approx& \frac{1}{\sqrt{2}}\left(1 -  \zeta - \frac{1}{2} \zeta^2\right)  \;.
	\end{eqnarray}
	It is then straightforward to obtain the nine matrix elements of $U$ by substituting Eq.~(\ref{eq:sin_cos_theta}) into Eq.~(\ref{eq:PMNS_standard}). The results are:
	\begin{eqnarray}
		\label{eq:U_alpha_i_vacuum}
		U_{e1}^{} &\approx& \frac{\sqrt{2} + 1}{3} - \frac{1}{6} \left[\left(2+\sqrt{2}\right)A + 2\left(\sqrt{2}-1\right)B\right] \zeta \nonumber \\
		&& - \frac{1}{6} \left[\left(\sqrt{2}+1\right)A^2 - \left(2-\sqrt{2}\right)A B + \left(\sqrt{2}+1\right)B^2\right] \zeta^2 \;, \nonumber \\
		U_{e2}^{} &\approx& \frac{\sqrt{2}+1}{3\sqrt{2}} + \frac{1}{6} \left[2\left(\sqrt{2} + 1\right)A - \left(2-\sqrt{2}\right)B\right] \zeta \nonumber \\
		&& - \frac{1}{12} \left[\left(2+\sqrt{2}\right)A^2 + 4\left(\sqrt{2}-1\right)AB + \left(2+\sqrt{2}\right) B^2\right] \zeta^2  \;, \nonumber \\
		U_{e3}^{} &\approx& \frac{\sqrt{2}-1}{\sqrt{6}}{\rm e}^{-{\rm i}\delta} + \frac{\left(\sqrt{2} + 1\right)B}{\sqrt{6}} \zeta {\rm e}^{-{\rm i}\delta} - \frac{\left(2-\sqrt{2}\right)B^2}{4\sqrt{3}} \zeta^2 {\rm e}^{-{\rm i}\delta} \;, \nonumber \\
		U_{\mu 1}^{} &\approx& -\frac{1}{\sqrt{6}} - \frac{\sqrt{2} -1}{3\sqrt{2}} {\rm e}^{{\rm i}\delta} + \frac{1}{6} \left\{\sqrt{6} - 2\sqrt{3}A - \left[2-\sqrt{2} - \left(\sqrt{2}-1\right) A + \left(2+\sqrt{2}\right) B\right] {\rm e}^{{\rm i} \delta}\right\} \zeta \nonumber \\
		&& + \frac{1}{12} \left\{\sqrt{6} + \sqrt{3}\left(\sqrt{2}A + 4\right) A + \left[2-\sqrt{2} + \left(2-\sqrt{2}\right) A^2 - 2 \left(2+\sqrt{2}\right)B  \right. \right. \nonumber \\
		&& \left. \left.  + \left(2-\sqrt{2}\right) B^2 + 2A\left(\sqrt{2}-1+B+\sqrt{2}B\right) \right]{\rm e}^{{\rm i}\delta}\right\} \zeta^2 \;, \nonumber \\
		U_{\mu 2}^{} &\approx& \frac{1}{\sqrt{3}} - \frac{\sqrt{2}-1}{6}{\rm e}^{{\rm i}\delta} - \frac{1}{6} \left\{2\sqrt{3} + \sqrt{6}A + \left[\sqrt{2}-1+\left(2-\sqrt{2}\right)A + \left(\sqrt{2}+1\right)B\right]{\rm e}^{{\rm i}\delta}\right\}  \zeta \nonumber \\
		&& - \frac{1}{12\sqrt{2}} \left\{2\sqrt{6} - 4\sqrt{3} A + 2\sqrt{6} A^2 - \left[2-\sqrt{2} + \left(2-\sqrt{2}\right)A^2 - 2 \left(2+\sqrt{2}\right)B \right.\right. \nonumber\\
		&& \left.\left. + \left(2-\sqrt{2}\right)B^2-4A\left(\sqrt{2}-1+B+\sqrt{2}B\right)\right]{\rm e}^{{\rm i}\delta}\right\} \zeta^2 \;, \nonumber \\
		U_{\mu 3}^{} &\approx& \frac{\sqrt{2} + 1}{2\sqrt{3}} + \frac{1+\sqrt{2}-\left(\sqrt{2}-1\right)B}{2\sqrt{3}} \zeta - \frac{1+\sqrt{2} + \left(2\sqrt{2}-2+B+\sqrt{2}B\right)B}{4\sqrt{3}} \zeta^2 \;, \nonumber \\
		U_{\tau 1}^{} &\approx&  \frac{1}{\sqrt{6}} - \frac{\sqrt{2} -1}{3\sqrt{2}} {\rm e}^{{\rm i}\delta} + \frac{1}{6} \left\{\sqrt{6} + 2\sqrt{3}A + \left[2-\sqrt{2} + \left(\sqrt{2}-1\right) A - \left(2+\sqrt{2}\right) B\right] {\rm e}^{{\rm i} \delta}\right\} \zeta \nonumber \\
		&& - \frac{1}{12} \left\{\sqrt{6} + \sqrt{3}\left(\sqrt{2}A - 4\right) A - \left[2-\sqrt{2} + \left(2-\sqrt{2}\right) A^2 + 2 \left(2+\sqrt{2}\right)B  \right. \right. \nonumber \\
		&& \left. \left.  + \left(2-\sqrt{2}\right) B^2 - 2A\left(\sqrt{2}-1-B-\sqrt{2}B\right) \right]{\rm e}^{{\rm i}\delta}\right\} \zeta^2 \;, \nonumber \\
		U_{\tau2}^{} &\approx& -\frac{1}{\sqrt{3}} - \frac{\sqrt{2}-1}{6}{\rm e}^{{\rm i}\delta} - \frac{1}{6} \left\{2\sqrt{3} - \sqrt{6}A - \left[\sqrt{2}-1-\left(2-\sqrt{2}\right)A - \left(\sqrt{2}+1\right)B\right]{\rm e}^{{\rm i}\delta}\right\} \zeta \nonumber \\
		&& + \frac{1}{12\sqrt{2}} \left\{2\sqrt{6} + 4\sqrt{3} A + 2\sqrt{6} A^2 + \left[2-\sqrt{2} + \left(2-\sqrt{2}\right)A^2 + 2 \left(2+\sqrt{2}\right)B \right.\right. \nonumber\\
		&& \left.\left. + \left(2-\sqrt{2}\right)B^2 + 4A\left(\sqrt{2}-1-B-\sqrt{2}B\right)\right]{\rm e}^{{\rm i}\delta}\right\} \zeta^2 \;, \nonumber \\
		U_{\tau 3}^{} &\approx& \frac{\sqrt{2} + 1}{2\sqrt{3}} - \frac{1+\sqrt{2}+\left(\sqrt{2}-1\right)B}{2\sqrt{3}} \zeta - \frac{1+\sqrt{2} - \left(2\sqrt{2}-2-B-\sqrt{2}B\right)B}{4\sqrt{3}} \zeta^2 \;, 
	\end{eqnarray}
	up to ${\cal O}\left(\zeta^3\right)$ or ${\cal O}\left(10^{-6} \right)$ in a good accuracy for present and future experiments. 
	
	With expressions of $U_{\alpha i}^{}$ above, some useful comments and discussions are in order.
	\begin{itemize}
		\item The $\mu$-$\tau$ flavor symmetry breaking caused by $\zeta$ can be described by defining the following three rephasing invariants:
	\begin{eqnarray}
		\label{eq:mu-tau-asymmetry}
		\Delta_1^{} &\equiv& \left|U_{\tau 1}^{}\right|^2 - \left|U_{\mu 1}^{}\right|^2 \nonumber \\
		&\approx& -\frac{2\left(\sqrt{2}-1\right)}{3\sqrt{3}}\cos\delta + \frac{4\sqrt{2} - \sqrt{3}\left[\left(2-\sqrt{2} \right)A+2\left(\sqrt{2}+1 \right)B  \right]\cos\delta}{9} \zeta \nonumber\\
		&& + \frac{2\left(9\sqrt{2}-4 \right)A-4B+\sqrt{3}\left[\left(\sqrt{2}-1\right)\left(4A^2+B^2+4\right) - \left(2+\sqrt{2}\right)A B \right] \cos\delta}{9} \zeta^2  \;, \nonumber \\
		\Delta_2^{} &\equiv& \left|U_{\tau 2}^{}\right|^2 - \left|U_{\mu 2}^{}\right|^2 \nonumber \\
		&\approx& +  \frac{2\left(\sqrt{2}-1\right)}{3\sqrt{3}}\cos\delta + \frac{9+2\sqrt{2}+\sqrt{3}\left[\left(2-\sqrt{2}\right)A+2\left(\sqrt{2}+1\right)B\right]\cos\delta}{9} \zeta \nonumber\\
		&& - \frac{2\left(9\sqrt{2}-4 \right)A+2B+\sqrt{3}\left[\left(\sqrt{2}-1\right)\left(4A^2+B^2+4\right)-\left(2+\sqrt{2}\right)A B \right] \cos\delta}{9} \zeta^2  \;, \nonumber \\
		\Delta_3^{} &\equiv& \left|U_{\tau 3}^{}\right|^2 - \left|U_{\mu 3}^{}\right|^2 \approx -\frac{3+2\sqrt{2}}{3}\zeta + \frac{2}{3}B\zeta^2  \;.
	\end{eqnarray}
	It is obvious that $\Delta_1^{}+ \Delta_2^{}+\Delta_3^{}=0$ holds due to the unitarity of the PMNS matrix. One can easily find that $\Delta_i^{}=0$ (for $i=1,2,3$) requires both $\zeta=0$ (or equivalently $\theta_{23}^{} = 45^\circ$) and $\delta=\pm\pi/2$, which are also the same conditions for the unitary PMNS matrix to keep the exact $\mu$-$\tau$ symmetry~\cite{Harrison:2002et,Fukuyama:1997ky}. The latest measurement of neutrino oscillation parameters presented jointly by Super-Kamiokande and T2K  constrains $\delta \approx -1.76$~\cite{T2K:2024wfn}. Therefore, the values of $\Delta_i^{}$ are all of ${\cal O}\left(10^{-2} \right)$, indicating the $\mu$-$\tau$ flavor symmetry could be treated as a more fundamental symmetry applied to the model building of leptonic mixing pattern.
	
	\item The Jarlskog invariant~\cite{Jarlskog:1985ht,Wu:1985ea} can be directly calculated through its definition:
	\begin{eqnarray}
		\label{eq:vacuum_J}
		{\cal J} \equiv {\rm Im}\left(U_{e2}^{} U_{\mu 3} U_{e3}^* U_{\mu2}^*\right) \approx \frac{\sqrt{2}+1}{36\sqrt{3}} \sin\delta + \frac{\left(2+\sqrt{2} \right)A + \left(18+6\sqrt{2} \right)B  }{72\sqrt{3}}\zeta\sin\delta \;.
	\end{eqnarray} 
	The first term agrees with the result in Ref.~\cite{Xing:2010pn}, which could be treated as the prediction of the NNE approach. With the numerical values of $\zeta$, $A$ and $B$ listed in Table~\ref{table:zeta_A_B_9.7}, we evaluate ${\cal J}\approx 3.38 \times 10^{-2}\sin\delta$, which is close to the experiment constraints from T2K and Super-Kamiokande with the maximum possibility~\cite{T2K:2024wfn}.
	
	\item The ordering of nine matrix elements deserves further discussions as it directly reflects the different mixing behavior between leptons and quarks. The smallness of $\zeta$ ensures that a reliable result can be directly given by the pattern of $U_0^{}$, which indicates $|U_{e1}^{}|>|U_{e2}^{}|>|U_{e3}^{}|$, $|U_{\mu 1}^{}|<|U_{\mu 2}^{}|<|U_{\mu 3}^{}|$ and $|U_{\tau 1}^{}|<|U_{\tau 2}^{}|<|U_{\tau 3}^{}|$. On the other hand, one may obtain $|U_{\mu i}^{}|\approx |U_{\tau i}^{}|$ from Eq.~(\ref{eq:mu-tau-asymmetry}) for $i=1,2,3$, and the specific value of $\zeta$ is needed to decide their ordering. Furthermore, one can easily obtain
	\begin{eqnarray}
		|U_{e1}^{}|^2-|U_{\mu 3}^{}|^2 \approx |U_{\mu 3}^{}|^2-|U_{e2}^{}|^2 \approx \frac{1}{36} \left(2 \sqrt{2}+3\right) >0 \;,
	\end{eqnarray}
	indicating $|U_{e1}^{}| > \{|U_{\mu 3}^{}|,|U_{\tau 3}^{}|\} > |U_{e2}^{}|$, and
	\begin{eqnarray}
		|U_{\mu 1}^{}|^2-|U_{e3}^{}|^2 \approx \frac{1}{18} \left[2 \left(\sqrt{2}-1\right) \sqrt{3} \cos\delta +4 \sqrt{2}-3\right] \;,
	\end{eqnarray}
	which is always positive for $\delta \in [-\pi,\pi]$. So far, we can determine the following ordering:
	\begin{eqnarray}
		|U_{e1}^{}| > \{|U_{\mu 3}^{}|,|U_{\tau 3}^{}|\} > \{|U_{e2}^{}|,|U_{\mu 2}^{}|,|U_{\tau 2}^{}|\} > \{|U_{\mu 1}^{}|,|U_{\tau 1}^{}|\} > |U_{e3}^{}| \;.
	\end{eqnarray}
	With the corresponding values of $\zeta$, $A$ and $B$ listed in Table~\ref{table:zeta_A_B_9.7} and the best-fit value of $\delta$ in Refs.~\cite{Gonzalez-Garcia:2021dve,Capozzi:2021fjo}, we have $\Delta_1^{} \approx 0.06\ (0.02)\times 10^{-2}$, $\Delta_2^{} \approx -0.2\ (+0.1)$ and $\Delta_3^{} \approx +0.1\ (-0.1)$ in the NO (IO) case. Therefore, we have 
	\begin{eqnarray}
		|U_{e1}^{}| > |U_{\tau 3}^{}| > |U_{\mu 3}^{}| > |U_{\mu 2}^{}| > |U_{e2}^{}| > |U_{\tau 2}^{}| > |U_{\tau 1}^{}| > |U_{\mu 1}^{}| > |U_{e3}^{}| \;,
	\end{eqnarray}
	for NO and 
	\begin{eqnarray}
		|U_{e1}^{}| > |U_{\mu 3}^{}| > |U_{\tau 3}^{}| > |U_{\tau 2}^{}| > |U_{\mu 2}^{}| > |U_{e2}^{}| > |U_{\tau 1}^{}| > |U_{\mu 1}^{}| > |U_{e3}^{}| \;,
	\end{eqnarray}
	for IO. The above results are also consistent with those obtained in Ref.~\cite{Xing:2024pal} since we both adopt the best-fit values.
	\end{itemize}
	
	Taking this opportunity, we would like to make a comparison between our strategy and the Wolfenstein-like expansion made in Ref.~\cite{Xing:2024pal}. Generally speaking, a small expansion parameter based on the constant flavor mixing matrix is both chosen to represent the mixing angles and matrix elements. However, in this work, we start from the experimental data, select the constant mixing matrix whose predicted mixing angles are closer to the best-fit values, and then gradually approach the true values through perturbation. Instead, the author of Ref.~\cite{Xing:2024pal} examines the special structure of the PMNS matrix by concentrating on the smallest matrix element. Therefore, the constant mixing matrices are chosen differently, even if the perturbative expansion is carried out in a comparable way.

	\section{The Unitarity Triangle}
	\label{sec:unitarity}
	
	The unitary PMNS matrix indicates two sets of relations among the matrix elements:
	\begin{eqnarray} \label{eq:unitarity_condition}
		\sum_i U_{\alpha i}^{} U_{\beta i}^* = \delta_{\alpha\beta}^{} \;, \quad \sum_\alpha U_{\alpha i}^{} U_{\alpha j}^* = \delta_{ij}^{} \;,
	\end{eqnarray}
	defining six so-called PMNS unitarity triangles $\left(\Delta_e^{},\Delta_\mu^{},\Delta_\tau^{} \right) $ and $\left(\Delta_1^{},\Delta_2^{},\Delta_3^{} \right) $ in the complex plane~\cite{Fritzsch:1999ee}. Such a geometric description is useful to analyze neutrino flavor conversions. In this work, we focus on the triangle $\Delta_\tau^{} $:
	\begin{eqnarray}
		\Delta_\tau^{}: \quad  U_{e1}^{} U_{\mu 1}^* + U_{e2}^{} U_{\mu 2}^* + U_{e3}^{} U_{\mu 3}^* = 0 \;,
	\end{eqnarray}
	or equivalently the {\it rescaled} triangle $\Delta_\tau'$:
	\begin{eqnarray}
		\Delta_\tau' :\quad \frac{U_{e1}^{} U_{\mu 1}^*}{U_{e3}^{} U_{\mu 3}^*} + \frac{U_{e2}^{} U_{\mu 2}^*}{U_{e3}^{} U_{\mu 3}^*} + 1 = 0 \;,
	\end{eqnarray}
	as it is directly related to the appearance channel $\nu_\mu^{} \to \nu_e^{}$ and its CP-conjugated process $\overline{\nu}_\mu^{} \to \overline{\nu}_e^{}$. These channels are crucial for next-generation long-baseline accelerator neutrino oscillation experiments, e.g., T2HK~\cite{Hyper-Kamiokande:2018ofw} and DUNE~\cite{DUNE:2015lol}, to probe the leptonic CP-violation. With Eq.~(\ref{eq:U_alpha_i_vacuum}), two sloping sides of $\Delta_\tau' $ read
	\begin{eqnarray}\label{eq:vacuum_sloping}
		\frac{U_{e1}^{} U_{\mu 1}^*}{U_{e3}^{} U_{\mu 3}^*} &\approx& -\frac{2}{3} - \frac{2\left(\sqrt{2} + 1\right)}{\sqrt{3}} {\rm e}^{{\rm i}\delta}  + \frac{2 \sqrt{2}}{3}  A \zeta -\frac{ \left(\sqrt{2}+2\right) A-2 \left(5 \sqrt{2} B+7 B+2 \sqrt{2}+2\right)}{\sqrt{3}} \zeta {\rm e}^{{\rm i} \delta } \;, \nonumber \\
		\frac{U_{e2}^{} U_{\mu 2}^*}{U_{e3}^{} U_{\mu 3}^*} &\approx& -\frac{1}{3} + \frac{2\left(\sqrt{2}+1\right)}{\sqrt{3}}   {\rm e}^{{\rm i}\delta}  -\frac{2 \sqrt{2}}{3}  A \zeta +\frac{ \left(\sqrt{2}+2\right) A-2 \left(5 \sqrt{2} B+7 B+2 \sqrt{2}+2\right)}{\sqrt{3}} \zeta {\rm e}^{{\rm i} \delta }  \;.
	\end{eqnarray}
	Strictly speaking, since the values of $\zeta$, $A$ and $B$ listed in Table~\ref{table:zeta_A_B_9.7} are dependent on the mass ordering, there will be certain differences in the shape of $\Delta_\tau'$ between the cases of NO and IO. Nevertheless, such dependence will appear only in higher orders, while the shape of $\Delta_\tau'$ is described by constant terms at the leading order, i.e., the first two terms in Eq.~(\ref{eq:vacuum_sloping}). To be more intuitive, choosing $\delta = -\pi/2$ as the benchmark, we arrive at $\left|U_{e1}^{} U_{\mu 1}^*\right|/\left|U_{e3}^{} U_{\mu 3}^*\right| \approx 3.38$ and $\left|U_{e2}^{} U_{\mu 2}^*\right|/\left|U_{e3}^{} U_{\mu 3}^*\right| \approx 3.32$, indicating two much longer sloping sides. With the help of Eq.~(\ref{eq:vacuum_J}), the height of the rescaled triangle can be expressed as
	\begin{eqnarray}
		\label{eq:height_vacuum_9.7}
		{\cal J}' = \frac{{\cal J}}{\left|U_{e3}^{} U_{\mu 3}^*\right|^2} \approx \frac{\left(\sqrt{2}+1\right)}{\sqrt{3}}\left\{ 2 - \left[ 4-\sqrt{2} A+\left(6+4\sqrt{2}\right)B\right] \zeta\right\} \sin\delta \;,
	\end{eqnarray}
	which is $\left|{\cal J}'\right| \approx 2.79$ for $\delta = -\pi/2$.
	
	In long-baseline accelerator neutrino experiments, terrestrial matter effects play an important role in analyzing neutrino flavor conversion and precisely measuring oscillation parameters. The shape of unitarity triangles could be modified, and thus affect the measurements of leptonic CP violation. In this work, all the effective parameters in matter are marked by a tilde. With the mixing matrix $\widetilde{U}$ in matter, the {\it effective} rescaled unitary triangle $\widetilde{\Delta}_\tau'$ can be drawn through
	\begin{eqnarray}
		\label{eq:rescaled_unitarity_matter}
		\widetilde{\Delta}_\tau' :\quad \frac{\widetilde{U}_{e1}^{} \widetilde{U}_{\mu 1}^*}{\widetilde{U}_{e3}^{} \widetilde{U}_{\mu 3}^*} + \frac{\widetilde{U}_{e2}^{} \widetilde{U}_{\mu 2}^*}{\widetilde{U}_{e3}^{} \widetilde{U}_{\mu 3}^*} + 1 = 0 \;.
	\end{eqnarray}
	In the following, we try to discuss the shape of $\widetilde{\Delta}_\tau' $ from three aspects. We will adopt (a) the exact analytical formulas of the effective parameters (see, e.g., Refs.~\cite{Barger:1980tf,Zaglauer:1988gz,Xing:2000gg}), (b) the approximation made in Ref.~\cite{Xing:2016ymg}, which applies to the neutrino with energy $E \lesssim 1~{\rm GeV}$, and (c) the so-called Freund's approximation~\cite{Freund:2001pn}, available for energy larger than $1~{\rm GeV}$, but still has a satisfying degree of accuracy for neutrino energy of ${\cal O}\left(0.1~{\rm GeV}\right)$ (see discussions in Ref.~\cite{Li:2016pzm}). Specifically speaking:
	\begin{enumerate}[(a)]
		\item {\bf Exact formulas}. One can always exactly express the effective parameters, such as neutrino masses $\widetilde{m}_i^{}$ (for $i=1,2,3$) and mixing matrix $\widetilde{U}$ (expressed with mixing angles $\widetilde{\theta}_{ij}^{}$ and CP phase $\widetilde{\delta}$), with those in vacuum and the matter parameter $a \equiv 2\sqrt{2}G_{\rm F}^{} N_e^{} E$. Here $G_{\rm F}^{} \approx 1.166 \times 10^{-5}~{\rm GeV}^{-2}$ is the Fermi constant, $N_e^{}$ and $E$ are electron number density and neutrino energy, respectively. Then the three terms in the left-handed side of Eq.~(\ref{eq:rescaled_unitarity_matter}) can be expressed with the help of~\cite{Xing:2003ez,Zhang:2004hf}
		\begin{eqnarray}
			\label{eq:exact_UUtilde}
			\widetilde{U}_{e 1}^{} \widetilde{U}_{\mu 1}^* &=& \frac{\widehat{\Delta}_{21}^{} \Delta_{31}^{}}{\widetilde{\Delta}_{21}^{} \widetilde{\Delta}_{31}^{}} U_{e 1}^{} U_{\mu 1}^* + \frac{\widehat{\Delta}_{11}^{} \Delta_{32}^{}}{\widetilde{\Delta}_{21}^{} \widetilde{\Delta}_{31}^{}} U_{e 2} U_{\mu 2}^* \;, \nonumber \\
			\widetilde{U}_{e 2}^{} \widetilde{U}_{\mu 2}^* &=& \frac{\widehat{\Delta}_{32}^{} \Delta_{21}^{}}{\widetilde{\Delta}_{32}^{} \widetilde{\Delta}_{21}^{}} U_{e 2}^{} U_{\mu 2}^* + \frac{\widehat{\Delta}_{22}^{} \Delta_{31}^{}}{\widetilde{\Delta}_{32}^{} \widetilde{\Delta}_{21}^{}} U_{e 3} U_{\mu 3}^* \;, \nonumber \\
			\widetilde{U}_{e 3}^{} \widetilde{U}_{\mu 3}^* &=& \frac{\widehat{\Delta}_{13}^{} \Delta_{23}^{}}{\widetilde{\Delta}_{13}^{} \widetilde{\Delta}_{23}^{}} U_{e 3}^{} U_{\mu 3}^* + \frac{\widehat{\Delta}_{33}^{} \Delta_{21}^{}}{\widetilde{\Delta}_{13}^{} \widetilde{\Delta}_{23}^{}} U_{e 1} U_{\mu 1}^* \;,
		\end{eqnarray}
		where $\Delta_{ji}^{} \equiv \Delta m_{ji}^2 = m_j^2 - m_i^2$, $\widehat{\Delta}_{ji}^{} \equiv m_j^2 - \widetilde{m}_i^2$ and $\widetilde{\Delta}_{ji}^{} \equiv \widetilde{m}_j^2 - \widetilde{m}_i^2$. The specific forms of $\widetilde{m}_i^2$ are listed in Appendix~\ref{app:A}, and the mixing matrix elements in vacuum $U_{\alpha i}^{}$ can be directly read off from Eq.~(\ref{eq:U_alpha_i_vacuum}).
		
		\item {\bf Low-energy approximation}. In such an approximation, two small expansion parameters $\alpha \equiv \Delta m_{21}^2/\Delta m_{31}^2$ and $\beta\equiv a/\Delta m_{31}^2$ are introduced. Then we arrive at $\widetilde{U}_{e3}^{} \widetilde{U}_{\mu 3}^* \approx U_{e3}^{} U_{\mu 3}^*$ and~\cite{Xing:2016ymg}
		\begin{eqnarray}\label{eq:matter_sloping_Xing_Zhu}
			\frac{\widetilde{U}_{e1}^{} \widetilde{U}_{\mu 1}^*}{\widetilde{U}_{e3}^{} \widetilde{U}_{\mu 3}^*} &\approx& \frac{\alpha}{\epsilon}\frac{U_{e1}^{} U_{\mu 1}^*}{U_{e3}^{} U_{\mu 3}^*} - \frac{\epsilon - \alpha - \beta}{2\epsilon}\;,  \nonumber \\
			\frac{\widetilde{U}_{e2}^{} \widetilde{U}_{\mu 2}^*}{\widetilde{U}_{e3}^{} \widetilde{U}_{\mu 3}^*} &\approx& \frac{\alpha}{\epsilon}\frac{U_{e2}^{} U_{\mu 2}^*}{U_{e3}^{} U_{\mu 3}^*} - \frac{\epsilon - \alpha + \beta}{2\epsilon} \;, 
		\end{eqnarray}
		with 
		\begin{eqnarray}
			\epsilon \equiv \sqrt{\alpha^2-2\left(\left|U_{e1}^{}\right|^2 - \left|U_{e2}^{}\right|^2\right)\alpha\beta + \left(1-\left|U_{e3}^{}\right|^2\right)^2\beta^2} \;.
		\end{eqnarray} 
		Note that Eq.~(\ref{eq:matter_sloping_Xing_Zhu}) is valid only for the NO case, and the replacement $\epsilon\to -\epsilon$ should be performed in the IO case. Substituting the matrix elements in Eq.~(\ref{eq:U_alpha_i_vacuum}) into Eq.~(\ref{eq:matter_sloping_Xing_Zhu}) and keeping the leading-order terms, we arrive at
		\begin{eqnarray}\label{eq:matter_sloping_NNE}
			\frac{\widetilde{U}_{e1}^{} \widetilde{U}_{\mu 1}^*}{\widetilde{U}_{e3}^{} \widetilde{U}_{\mu 3}^*} &\approx& \mp \left[\frac{1}{6}+\frac{2\left(\sqrt{2}+1\right)}{\sqrt{3}}{\rm e}^{{\rm i}\delta}\right]\frac{\alpha}{\epsilon}-\frac{1}{2}\left(1\mp\frac{\beta}{\epsilon}\right)\;,  \nonumber \\
			\frac{\widetilde{U}_{e2}^{} \widetilde{U}_{\mu 2}^*}{\widetilde{U}_{e3}^{} \widetilde{U}_{\mu 3}^*} &\approx& \pm \left[\frac{1}{6}+\frac{2\left(\sqrt{2}+1\right)}{\sqrt{3}}{\rm e}^{{\rm i}\delta}\right]\frac{\alpha}{\epsilon}-\frac{1}{2}\left(1\pm\frac{\beta}{\epsilon}\right) \;, 
		\end{eqnarray}
		where the upper (lower) sign is for the NO (IO) case. With the approximation $\widetilde{\cal J}/{\cal J} \approx |\alpha|/\epsilon$ derived in Ref.~\cite{Xing:2016ymg}, where $\widetilde{\cal J}$ represents the effective Jarlskog invariant in matter, the height of $\widetilde{\Delta}_\tau'$ is
		\begin{eqnarray}
			\label{eq:low_energy_height}
			\widetilde{\cal J}' = \frac{\widetilde{\cal J}}{\left|\widetilde{U}_{e3}^{} \widetilde{U}_{\mu3}^*\right|^2} \approx \frac{\left(\sqrt{2}+1\right)\alpha}{\sqrt{3}\epsilon} \left\{2 - \left[ 4-\sqrt{2} A+\left(6+4\sqrt{2}\right)B\right] \zeta\right\} \sin\delta\;.
		\end{eqnarray}
		Since $\alpha/\epsilon<1$, the height of $\widetilde{\Delta}_\tau'$ is shorter than the vacuum one in Eq.~(\ref{eq:height_vacuum_9.7}).
		
		\item {\bf Freund's approximation}. The formulas in Freund's seminal paper can be adopted to calculate the effective mixing angles in matter $\widetilde{\theta}_{ij}^{}$ with higher neutrino energies. With $\alpha$ and $\beta$ defined above and $C \equiv \sqrt{\beta^2 -2\beta \cos 2 \theta_{13}^{} +1} $, one obtains~\cite{Freund:2001pn} 
		\begin{eqnarray}
			\sin2\widetilde{\theta}_{12}^{} &=&  \frac{2\alpha C \sin 2\theta_{12}^{}}{\left|\beta\right| \cos\theta_{13}^{} \sqrt{2C\left(-\beta+C+\cos 2\theta_{13}^{} \right) }} \;, \nonumber \\
			\sin^2 2\widetilde{\theta}_{13}^{} &=& \frac{\sin^2 2\theta_{13}^{}}{C^2} + \frac{2\alpha \beta \left(-\beta + \cos 2\theta_{13}^{}\right)\sin^2\theta_{12}^{} \sin^2 2 \theta_{13}^{}}{C^4} \;, \nonumber \\
			\sin2\widetilde{\theta}_{23}^{} &=& \sin2\theta_{23}^{} +   \frac{2\alpha\beta \sin 2\theta_{12}^{}\sin\theta_{13}^{}\cos2\theta_{23}^{}}{1+C-\beta\cos 2\theta_{13}^{}} \cos\delta  \;, \nonumber \\
			\sin\widetilde{\delta} &=& \sin\delta \left(1- \frac{\alpha\cos\delta}{\tan 2 \theta_{23}^{}} \frac{2\beta \sin 2\theta_{12}^{} \sin\theta_{13}^{}}{1+C-\beta\cos 2\theta_{13}^{}}\right) \;.
		\end{eqnarray}
		The $\hat{A}$ in Ref.~\cite{Freund:2001pn} is equal to $\beta$ by definition, and the upper sign in Eq.~(28) of Ref.~\cite{Freund:2001pn} is chosen since $\cos 2 \theta_{23}^{} < \beta $. Taking the effective mixing angles into the effective matrix elements, we obtain
		\begin{eqnarray}
			\widetilde{U}_{e1}^{} &\approx& \pm \frac{2\left(\sqrt{2}-1 \right) \alpha \sqrt{18C\left(2\sqrt{2} + 3C-3\beta \right) -3 }}{3|\beta|\left( 2\sqrt{2} + 3C-3\beta\right) } \;, \nonumber \\
			\widetilde{U}_{e2}^{} &\approx& \sqrt{1-\frac{1}{6C\left(2\sqrt{2}-3\beta+3C\right)}} \;, \nonumber \\
			\widetilde{U}_{e3}^{} &\approx& \frac{{\rm e}^{-{\rm i}\delta}}{\sqrt{6C\left(2\sqrt{2} + 3C-3\beta \right) }} \;, \nonumber \\
			\widetilde{U}_{\mu 1}^{} &\approx& - \frac{1}{\sqrt{2}} -  \frac{\left(2-\sqrt{2} \right)\alpha }{\sqrt{3}}\left[\pm\frac{{\rm e}^{{\rm i}\delta}}{|\beta|\left(2\sqrt{2} + 3C-3\beta \right)} - \frac{\beta\cos\delta}{3+3C-2\sqrt{2}\beta}\right] \;, \nonumber \\
			\widetilde{U}_{\mu 2}^{} &\approx& -  \frac{{\rm e}^{{\rm i}\delta}}{2\sqrt{3C\left(2\sqrt{2} + 3C-3\beta \right) }} \;, \nonumber \\
			\widetilde{U}_{\mu 3}^{} &\approx& \frac{1}{2}\sqrt{2-\frac{1}{6 \sqrt{2}C+9C^2-9C\beta}} \;, \nonumber \\
			\widetilde{U}_{\tau 1}^{} &\approx&  \frac{1}{\sqrt{2}} -  \frac{\left(2-\sqrt{2} \right)\alpha }{\sqrt{3}}\left[ \pm \frac{{\rm e}^{{\rm i}\delta}}{|\beta|\left(2\sqrt{2} + 3C-3\beta \right)} - \frac{\beta\cos\delta}{3+3C-2\sqrt{2}\beta}\right] \;, \nonumber \\
			\widetilde{U}_{\tau 2}^{} &\approx& -  \frac{{\rm e}^{{\rm i}\delta}}{2\sqrt{3C\left(2\sqrt{2} + 3C-3\beta \right) }} \;, \nonumber \\
			\widetilde{U}_{\tau 3}^{} &\approx& \frac{1}{2}\sqrt{2-\frac{1}{6 \sqrt{2}C+9C^2-9C\beta}} \;,
		\end{eqnarray}
		where the upper (lower) sign is for the NO (IO) case.\footnote{The dependence on the mass ordering comes from the fact that $\alpha$ reverses sign in the NO and IO cases, and we require $\sin^2\widetilde{\theta}_{12}^{}$ to increase with the neutrino energy~\cite{Li:2016pzm}.} We expand to the order of ${\cal O}(\alpha)$ for $\widetilde{U}_{\mu 2}^{}$ and $\widetilde{U}_{\tau 2}^{}$ to keep the Dirac CP-violation phase. The first six formulas are used to calculate $\widetilde{\Delta}_\tau'$ with two sloping sides
		\begin{eqnarray}\label{eq:matter_sloping_NNE_freund}
			\frac{\widetilde{U}_{e1}^{} \widetilde{U}_{\mu 1}^*}{\widetilde{U}_{e3}^{} \widetilde{U}_{\mu 3}^*} &\approx&  \mp \frac{4\sqrt{3}\left(\sqrt{2}-1 \right)\alpha C  }{|\beta|} {\rm e}^{{\rm i}\delta}\;, \nonumber \\ \frac{\widetilde{U}_{e2}^{} \widetilde{U}_{\mu 2}^*}{\widetilde{U}_{e3}^{} \widetilde{U}_{\mu 3}^*} &\approx& -1 \pm  \frac{4\sqrt{3}\left(\sqrt{2}-1 \right) \alpha C }{|\beta|}{\rm e}^{{\rm i}\delta}  \;, 
		\end{eqnarray}
		and the height
		\begin{eqnarray}
			\label{eq:freund_height}
			\widetilde{\cal J}' = \pm \frac{2\left(\sqrt{2} + 1 \right) }{\sqrt{3}} \sin\delta \times \frac{6 \left(3-2 \sqrt{2}\right) \alpha  C }{| \beta | }  \;.
		\end{eqnarray}
	\end{enumerate}
	
	\begin{table}[t]
		\centering
		\begin{spacing}{1.25}
			\begin{tabular}{ccccccc}
				\hline\hline
				&    & $a/\left(10^{-4}~{\rm eV}^2\right) $ & $\alpha/10^{-2}$ & $\beta/10^{-2}$ & $\epsilon/10^{-2}$ & $C$ \\ \hline
				\multirow{2}{*}{T2HK} & NO & \multirow{2}{*}{$ 1.18 $} & $+2.96$ & $+4.70$ & $4.38 $ & $0.96$ \\
				& IO &  & $-3.07$ & $-4.89$ & $4.56 $ & $1.05$ \\
				\multirow{2}{*}{DUNE} & NO & \multirow{2}{*}{$ 6.47 $} & $+2.96$ & $+25.8$ & $24.2$ & $0.76$ \\
				& IO & & $-3.07$  & $-26.8$ & $25.1$ & $1.26$  \\ \hline\hline
			\end{tabular}
		\end{spacing}
		\caption{The values of $\alpha$, $\beta$, $\epsilon$ and $C$ which are defined to describe matter effect for T2HK and DUNE in different cases of mass ordering. The best-fit values in Refs.~\cite{Capozzi:2021fjo,Gonzalez-Garcia:2021dve} are adopted. Neutrino energy $E=0.6\ (3.0)~{\rm GeV}$ and matter density $\rho=2.60\ (2.85)~{\rm g}/{\rm cm}^3$ are set as benchmark parameters for T2HK (DUNE). The electron fraction is assumed to be $Y_e^{} = 0.5$.}
		\label{table:T2HK_DUNE}
	\end{table}

	\begin{figure}[t]
		\centering
		\includegraphics[scale=0.37]{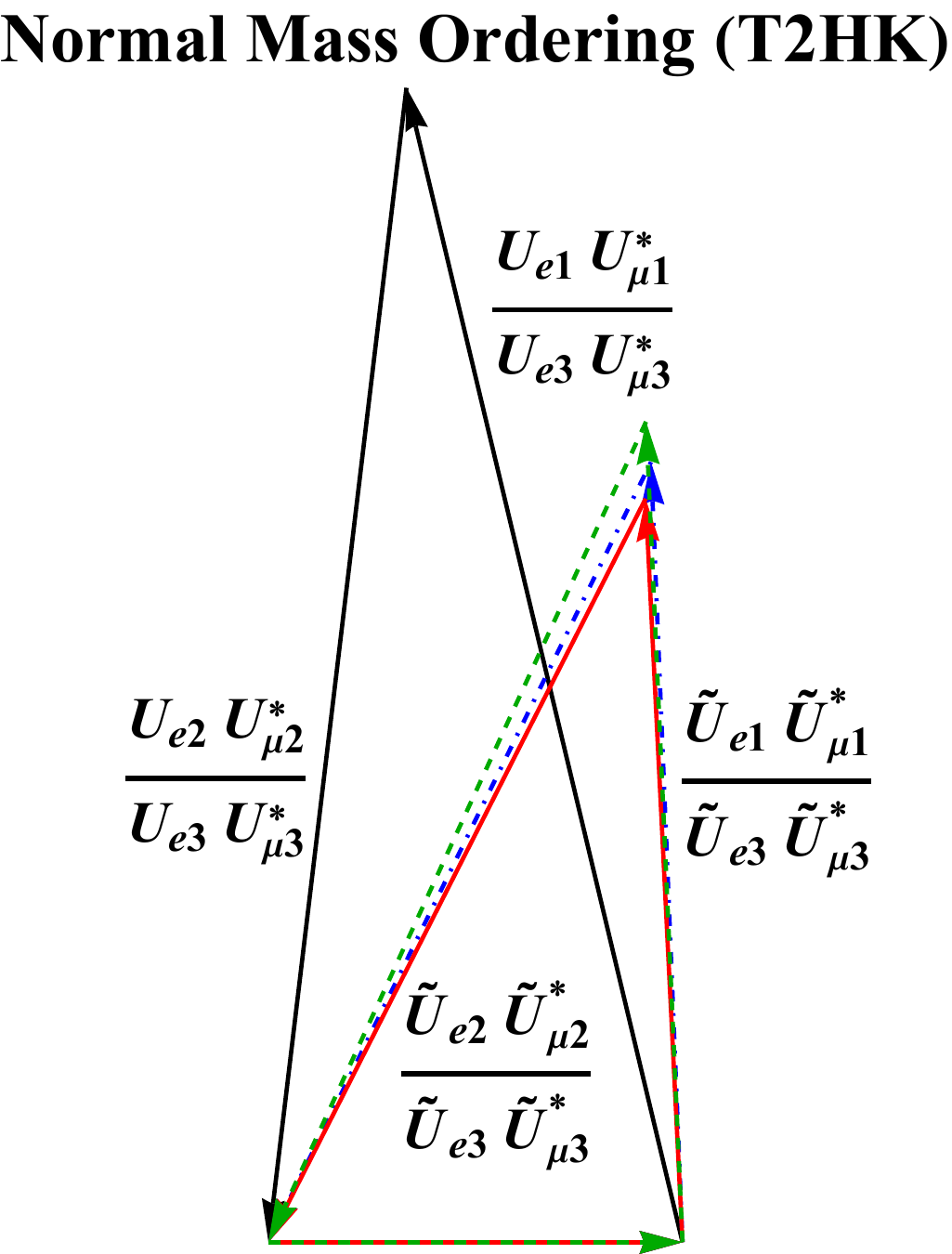}\qquad\qquad
		\includegraphics[scale=0.37]{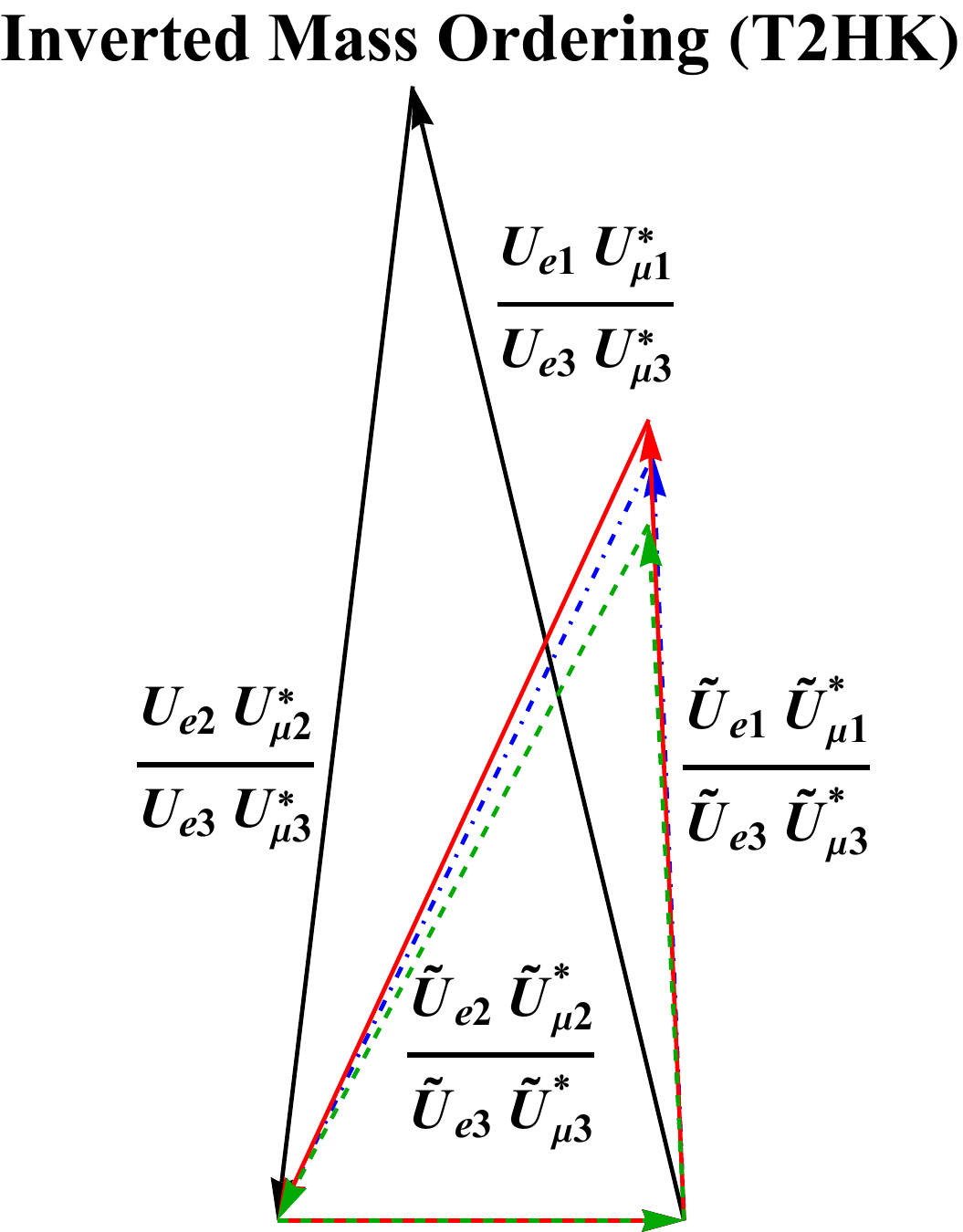}\\
		\caption{The unitarity triangles for T2HK in the NO (left) and IO (right) case. The benchmark values of neutrino energy and matter density are consistent with those in Table~\ref{table:T2HK_DUNE}, from which other parameters are also read off. The solid black triangle illustrates the vacuum one $\Delta_\tau'$, while the counterparts in matter $\widetilde{\Delta}_\tau'$ are plotted under the exact expressions from Refs.~\cite{Barger:1980tf,Zaglauer:1988gz,Xing:2000gg} (red solid), the approximation from Ref.~\cite{Xing:2016ymg} (blue dot-dashed), and the approximation with Freund's formulas~\cite{Freund:2001pn} (green dashed).}
		\label{fig:t2hk_UT}
	\end{figure}

	\begin{figure}[t]
		\centering
		\includegraphics[scale=0.37]{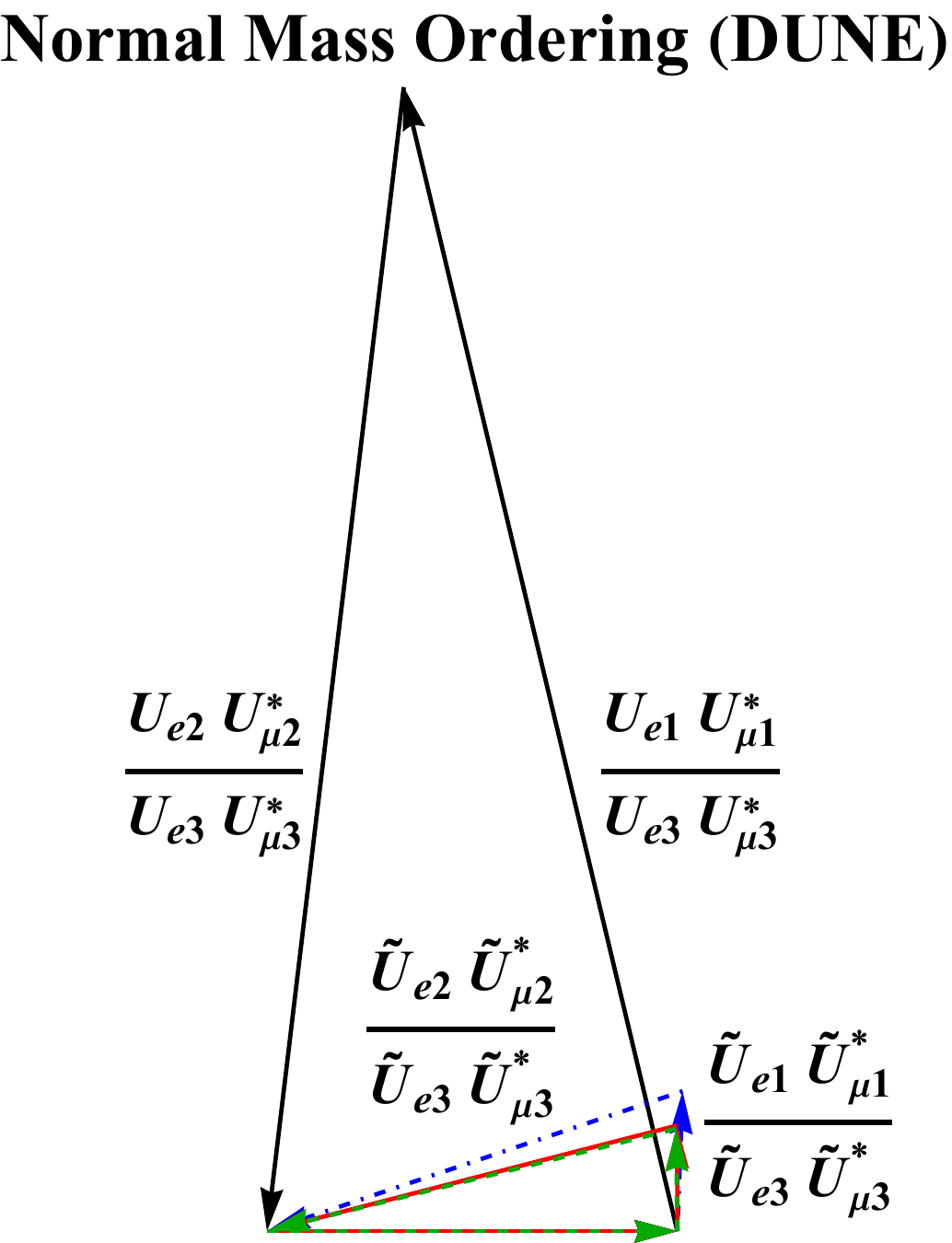}\qquad\qquad
		\includegraphics[scale=0.37]{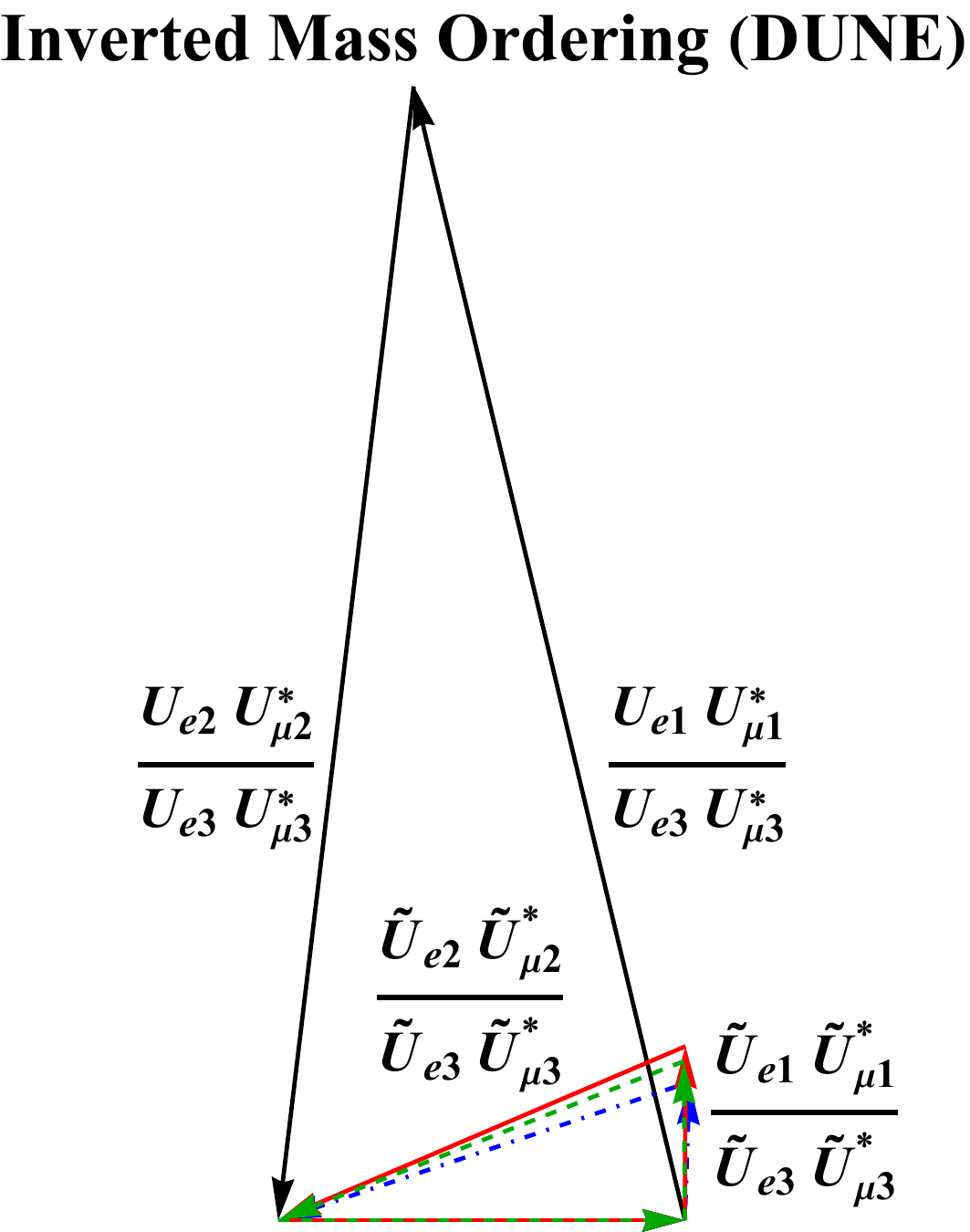}
		\caption{The unitarity triangles $\widetilde{\Delta}_\tau^{}$ and its counterpart $\widetilde{\Delta}_\tau'$ for DUNE, with the same input parameters and conventions in Fig.~\ref{fig:t2hk_UT}.}
		\label{fig:dune_UT}
	\end{figure}

	The energy spectrum of the neutrino beam produced in T2HK is peaked at $E\sim 0.6~{\rm GeV}$, and the Earth's density is $\rho\approx 2.60~{\rm g}/{\rm cm}^3$~\cite{Hyper-Kamiokande:2018ofw}, while for DUNE the typical energy can be as large as a few GeV with the matter density $\rho \approx 2.85~{\rm g}/{\rm cm}^3$~\cite{DUNE:2015lol}. Together with the best-fit values of three mixing angles and two mass-squared differences in Refs.~\cite{Capozzi:2021fjo,Gonzalez-Garcia:2021dve}, parameters related to matter effects, i.e., $a$, $\alpha$, $\beta$, $\epsilon$ and $C$, are evaluated in Table~\ref{table:T2HK_DUNE} for T2HK and DUNE in both NO and IO cases. We set $E=0.6\ (3.0)~{\rm GeV}$ as the benchmark energy of T2HK (DUNE) and assume the electron fraction $Y_e^{} = 0.5$. For antineutrinos, the matter parameter $a$ will flip its sign, and thus the values of $\beta$, $\epsilon$ and $C$ are also affected. With the help of Table~\ref{table:T2HK_DUNE}, we illustrate $\Delta_\tau'$ and its counterpart $\widetilde{\Delta}_\tau'$ in matter for T2HK and DUNE in Fig.~\ref{fig:t2hk_UT} and Fig.~\ref{fig:dune_UT}, respectively. Some useful comments are as follows:
	\begin{itemize}
		\item It is clear that two sloping sides of $\Delta_\tau'$ (solid black triangles) are much longer than the shortest one. However, the shape of $\widetilde{\Delta}_\tau'$ changes dramatically. We plot $\widetilde{\Delta}_\tau'$ in three colored triangles, corresponding to the three aspects discussed above. The red solid triangles are plotted by numerically solving the exact expressions in Eq.~(\ref{eq:exact_UUtilde}), while the blue dot-dashed and green dashed ones are respectively drawn from the low-energy approximation and Freund's approximation. The areas and heights of $\widetilde{\Delta}_\tau'$ are smaller than those of $\Delta_\tau'$, as indicated in Eqs.~(\ref{eq:low_energy_height}) and (\ref{eq:freund_height}). Meanwhile, as the beam energy of DUNE is larger, the height of $\widetilde{\Delta}_\tau'$ for DUNE is smaller than that of T2HK, which is consistent with the behavior of the Jarlskog invariant in matter~\cite{Zhang:2004hf}.
		
		\item In Fig.~\ref{fig:t2hk_UT}, the triangles plotted with the low-energy approximation are closer to the exactly solved ones, while such an approximation does not agree well for DUNE in Fig.~\ref{fig:dune_UT} since the neutrino beam energy is beyond the allowed range of the approximation. Although Freund's formulas do not fit well in the low-energy case, the differences are acceptable; when adopted on DUNE, they show excellent accuracy. Especially in the NO case (left panel of Fig.~\ref{fig:dune_UT}), the two triangles (red solid and green dashed) almost overlap with each other. 
	\end{itemize}
	Note that the shapes of $\widetilde{\Delta}_\tau'$ are not remarkably affected by the mass ordering. This can be easily understood from Eqs.~(\ref{eq:matter_sloping_NNE}) and (\ref{eq:matter_sloping_NNE_freund}). Taking the former one as an example. One may notice the ratios $\alpha/\epsilon$ and $\beta/\epsilon$ are always positive since the performance $\epsilon\to -\epsilon$ from NO to IO compensates the flipping sign of $\alpha$ and $\beta$. Besides, there is not much difference between the absolute values of $\alpha$, $\beta$ and thus $\epsilon$ in two cases of mass ordering (see Table.~\ref{table:T2HK_DUNE}). Therefore, two sloping sides in $\widetilde{\Delta}_\tau'$ will not change significantly.
	
	At the end of this section, we would like to mention that the unitarity conditions in Eq.~(\ref{eq:unitarity_condition}) may be violated in some circumstances, such as in the type-I or type-III seesaw model~\cite{Minkowski:1977sc,Yanagida:1979as,Gell-Mann:1979vob,Glashow:1979nm,Mohapatra:1979ia,Foot:1988aq}. In this case, the non-unitary leptonic mixing matrix $N$ can be given by $N = (\mathbbm{1}-\eta) U$, with the unitary matrix $U$ in Eq.~(\ref{eq:PMNS_standard}) and a Hermitian matrix $\eta$ characterized the unitarity derivations. The relation among matrix elements of $N$ can be expressed similarly to that in Eq.~(\ref{eq:unitarity_condition}):
	\begin{eqnarray}\label{eq:non_unitarity}
		\sum_i N_{\alpha i}^{} N_{\beta i}^{*} = \delta^{}_{\alpha \beta} - 2 \eta_{\alpha \beta}^{} + {\cal O}(\eta^2) \;,
	\end{eqnarray}
	which returns to the standard case when $\eta \to 0$. Now the unitarity triangle $\Delta_\tau^{}$ is modified to a quadrangle up to ${\cal O}(\eta^2)$:
	\begin{eqnarray}
		\Delta_\tau^{\rm n} : \quad N_{e 1}^{} N_{\mu 1}^{*} + N_{e 2}^{} N_{\mu 2}^{*} + N_{e 3}^{} N_{\mu 3}^{*} + 2 \eta_{e \mu}^{} = 0 \;.
	\end{eqnarray}
	One may examine the effects of the non-unitarity by studying the departure of the shape or the area of $\Delta_\tau^{\rm n}$ from $\Delta_\tau^{}$. The authors of Ref.~\cite{Luo:2023xmv} propose that a Pythagoras-like theorem in neutrino oscillation can be experimentally tested to investigate the non-unitarity of the mixing matrix. In a specific neutrino mass model, such as the type-I seesaw mechanism, with the given parametrization of the $6 \times 6$ mixing matrix, the non-unitary effect can be linked to the mixing angles between active and sterile neutrinos and their corresponding CP-violating phases~\cite{Xing:2011ur}. Finally, although the non-unitary effects of leptonic flavor mixing have been extensively studied, the latest global-fit results on the unitarity deviations constrain all matrix elements of $\eta$ to be between ${\cal O}(10^{-5})$ and ${\cal O}(10^{-3})$ at 95\% confidence level~\cite{Blennow:2023mqx}. Therefore, the discussions in this work are still reasonable even if we ignore such small deviations.
	
	\section{Summary}
	\label{sec:summary}
	
	Inspired by the well-studied {\it nearest-neighbor interaction} (NNI) in condensed matter physics, we put forward the NNE approach for research on leptonic flavor mixing. Similar to the dominant NNI between molecules in the lattice, the results obtained by our NNE can perfectly describe the neutrino oscillation behavior at the leading order. Starting with a chosen constant mixing pattern $U_0^{}$, we expand the mixing angles and matrix elements according to the expansion parameter $\zeta$. We study the $\mu$-$\tau$ permutation symmetry breaking, the leptonic CP violation, and the ordering of nine matrix elements. The unitarity triangle $\Delta_\tau'$ and its counterpart in matter $\widetilde{\Delta}_\tau'$ are plotted and the non-unitarity deviation effect is briefly analyzed. Compared with previous studies on $U$ with three expansion parameters for three mixing angles (e.g., works in Refs.~\cite{King:2007pr,Pakvasa:2007zj}), the proposed NNE approach can greatly simplify the calculations by introducing only one parameter.
	
	In the near future, the JUNO experiment will be devoted to determining the neutrino mass ordering and precisely measuring oscillation parameters to the sub-percent level~\cite{JUNO:2015zny,JUNO:2022mxj}, while T2HK and DUNE are expected to measure the CP-violating phase with high accuracy. Our NNE approach is valid for a wide range of constant mixing patterns, as long as the predicting mixing angles are close to the measurements. However, we should also note that there are still certain challenges in the model building, especially for those patterns with a nonzero $\theta_{13}^{(0)}$~\cite{King:2017guk,Feruglio:2019ybq,Xing:2020ijf,Ding:2023htn,Ding:2024ozt}. Nevertheless, the NNE can still be treated as a new perspective to research on neutrino phenomenology, such as the renormalization-group-equation (RGE) running effect and the flavor distribution of ultrahigh-energy (UHE) neutrinos.

	\section*{Acknowledgements}
	
	The author is indebted to Prof. Zhi-zhong Xing for valuable suggestions and partial involvement at the early stage of this work, and to Prof. Shun Zhou for carefully reading the manuscript. The author also thanks Dr. Jing-yu Zhu for useful discussions. This work was supported by the National Natural Science Foundation of China under grant No.~11835013 and the CAS Project for Young Scientists in Basic Research (YSBR-099).
	
	\appendix
	
	\section{Expressions of Effective Parameters}
	\label{app:A}
	
	\setcounterpageref{equation}{0}
	
	\numberwithin{equation}{section}
	
	The exact analytical expressions of effective neutrino mass squares $\widetilde{m}_i^2$ read~\cite{Barger:1980tf,Zaglauer:1988gz,Xing:2000gg}
	\begin{eqnarray}
		\widetilde{m}_1^2 &=& m_1^2 + \frac{1}{3} x-\frac{1}{3} \sqrt{x^2-3 y} \left[z+\sqrt{3\left(1-z^2\right)}\right] \;, \nonumber \\
		\widetilde{m}_2^2 &=& m_1^2+\frac{1}{3} x-\frac{1}{3} \sqrt{x^2-3 y}\left[z-\sqrt{3\left(1-z^2\right)}\right] \;,  \nonumber \\
		\widetilde{m}_3^2 &=& m_1^2+\frac{1}{3} x+\frac{2}{3} z \sqrt{x^2-3 y} \;, 
	\end{eqnarray}
	in the case of NO ($m_1^{} < m_2^{} < m_3^{}$), and 
	\begin{eqnarray}
		\widetilde{m}_1^2 &=& m_1^2 + \frac{1}{3} x-\frac{1}{3} \sqrt{x^2-3 y} \left[z-\sqrt{3\left(1-z^2\right)}\right] \;, \nonumber \\
		\widetilde{m}_2^2 &=& m_1^2+\frac{1}{3} x+\frac{2}{3} z \sqrt{x^2-3 y} \;,  \nonumber \\
		\widetilde{m}_3^2 &=& m_1^2+\frac{1}{3} x-\frac{1}{3} \sqrt{x^2-3 y}\left[z+\sqrt{3\left(1-z^2\right)}\right] \;, 
	\end{eqnarray}
	in the case of IO ($m_3^{} < m_1^{} < m_2^{}$), where 
	\begin{eqnarray}
		x &=& \Delta m_{21}^2 + \Delta m_{31}^2 + a \;, \nonumber \\
		y &=& \Delta m_{21}^2 \Delta m_{31}^2 + a \left[\Delta m_{21}^2 \left(1-\left|U_{e2}^{}\right|^2 \right) + \Delta m_{31}^2 \left(1-\left|U_{e3}^{}\right|^2 \right)  \right] \;, \nonumber \\
		z &=& \cos\left[\frac{1}{3} \arccos \frac{2x^3-9xy+27a \Delta m_{21}^2 \Delta m_{31}^2 \left|U_{e1}^{}\right|^2}{2\left(x^2-3y \right)^{3/2} }\right] \;.
	\end{eqnarray}
	Here the matter parameter $a \equiv 2\sqrt{2}G_{\rm F}^{} N_e^{} E$ and mass-squared difference $\Delta m_{ji}^2 \equiv m_j^2 - m_i^2$ are defined as usual.
	
	To obtain the effective mixing angles $\widetilde{\theta}_{ij}^{}$, we may use the relation
	\begin{eqnarray}
		\sin^2\widetilde{\theta}_{13}^{} = |\widetilde{U}_{e3}^{}|^2 \;, \quad \sin^2\widetilde{\theta}_{12}^{} = \frac{|\widetilde{U}_{e2}^{}|^2}{1-|\widetilde{U}_{e3}^{}|^2} \;, \quad \sin^2\widetilde{\theta}_{23}^{} = \frac{|\widetilde{U}_{\mu 3}^{}|^2}{1-|\widetilde{U}_{e3}^{}|^2} \;,
	\end{eqnarray}
	with~\cite{Zhang:2004hf}
	\begin{eqnarray}
		|\tilde{U}_{e 2}^{}|^2 &=&\frac{\widehat{\Delta}_{22}^{} \widehat{\Delta}_{32}^{}}{\widetilde{\Delta}_{12}^{} \widetilde{\Delta}_{32}^{}}\left|U_{e 1}^{}\right|^2+\frac{\widehat{\Delta}_{12}^{} \widehat{\Delta}_{32}^{}}{\widetilde{\Delta}_{12}^{} \widetilde{\Delta}^{}_{32}}\left|U^{}_{e 2}\right|^2+\frac{\widehat{\Delta}_{12}^{} \widehat{\Delta}^{}_{22}}{\widetilde{\Delta}^{}_{12} \widetilde{\Delta}^{}_{32}}\left|U^{}_{e 3}\right|^2 \;, \nonumber \\
		|\widetilde{U}_{e3}^{}|^2 &=&\frac{\widehat{\Delta}^{}_{23} \widehat{\Delta}^{}_{33}}{\widetilde{\Delta}_{13}^{} \widetilde{\Delta}^{}_{23}}\left|U^{}_{e 1}\right|^2+\frac{\widehat{\Delta}^{}_{13} \widehat{\Delta}^{}_{33}}{\widetilde{\Delta}_{13}^{} \widetilde{\Delta}^{}_{23}}\left|U^{}_{e 2}\right|^2+\frac{\widehat{\Delta}^{}_{13} \widehat{\Delta}^{}_{23}}{\widetilde{\Delta}_{13}^{} \widetilde{\Delta}^{}_{23}}\left|U^{}_{e 3}\right|^2 \;,  \nonumber \\
		|\widetilde{U}_{\mu 3}^{}|^2 &=&\frac{\widehat{\Delta}^{}_{11} \widehat{\Delta}^{}_{12}}{\widetilde{\Delta}_{31}^{} \widetilde{\Delta}^{}_{32}}\left|U^{}_{\mu  1}\right|^2 + \frac{\widehat{\Delta}^{}_{21} \widehat{\Delta}^{}_{22}}{\widetilde{\Delta}_{31}^{} \widetilde{\Delta}^{}_{32}}\left|U^{}_{\mu  2}\right|^2+\frac{\widehat{\Delta}^{}_{31} \widehat{\Delta}^{}_{32}}{\widetilde{\Delta}_{31}^{} \widetilde{\Delta}^{}_{32}}\left|U^{}_{\mu 3}\right|^2 \;.
	\end{eqnarray}
	We define $\widehat{\Delta}_{ji}^{} \equiv m_j^2 - \widetilde{m}_i^2$ and $\widetilde{\Delta}_{ji}^{} \equiv \widetilde{m}_j^2 - \widetilde{m}_i^2$.

\end{document}